\documentclass[a4paper]{jpconf}
\usepackage{graphicx}
\usepackage{xcolor}
\newcommand{\farcs}{\mbox{\ensuremath{.\!\!^{\prime\prime}}}}% 

\begin{document}
\title{Seeing measurements at OAUNI on 2016 and 2017 campaigns}

\author{A Pereyra$^1$$^,$$^2$, J Tello$^2$ and M Zevallos$^2$}

\address{$^1$ Geophysical Institute of Peru, Astronomy Area, Badajoz 169, Ate, Lima, Per\'u}
\address{$^2$ Faculty of Science, National University of Enginnering, Av. T\'upac Amaru 210, Lima, Per\'u}

\ead{apereyra@igp.gob.pe}

\begin{abstract}
We present seeing measurements at OAUNI site gathered on 2016 and 2017 campaigns using \textit{V} and \textit{R} broadband filters. In order to quantify the seeing we used the full-width-at-half-maximum from stellar profiles on photometric sequences during the observational windows of our supernovae program. A typical median seeing of 1$\farcs$8 was found on 2016 and a worst value of 2$\farcs$0 on 2017. The last one was probably affected by anomalous conditions related to the 2017 extreme climatic phenomena. The monthly first quartile analysis indicates that best seeing conditions can be achieved at a level of 1$\farcs$5. In general, our results indicate a reasonable sky quality for the OAUNI site.

\end{abstract}

\section{Introduction\label{intro}}

Site testing is crucial to optimize the selection of a given site with potential for astronomical observations. For this purpose is necessary a good knowledge of local meteorological factors which alter the quality of the astronomical images. In general, the level of atmospheric turbulence or seeing, the rate of clear skies and the luminosity level for the background sky are parameters to consider in this process. 

In particular, the deformations of the front wave from celestial bodies induced by changes on the atmospheric refraction index of turbulence layers \cite{tat61,fri65} characterizes the seeing disk measured on the focal plane of the optical system. As more atmospheric turbulence is present, the stellar profiles show bigger seeing disks and vice versa. The seeing can be measured in several ways, including direct measurements of the full-width-at-half-maximum (\textit{fwhm}) from stellar profiles or using special techniques as the differential image motion monitor \cite{sar90}. It can be demonstrated that the \textit{fwhm} of a long exposure stellar image measured at zenith angle (\textit{z}) and a given wavelength ($\lambda$) is proportional to sec(z)$^{0.6}\times\lambda^{-0.2}$ where sec(z) is the airmass ($\chi$) \cite{rod99}. In this computation the approximation of a plane-parallel atmosphere is assumed which implies zenith angles lower than 60$^{\circ}$ (or $\chi$ $<$ 2). These corrections have to be considered when targets, observed to different line of sight and filters, are used for comparative seeing studies. In particular, best seeing conditions are expected at longer wavelengths.

This work presents a systematic study of the astronomical seeing as part of a site testing program developed at the astronomical observatory of the National University of Enginnering \cite{per15} (OAUNI in spanish) on 2016 and 2017. This facility operates since 2015 at the Huancayo site of the Geophysical Institute of Peru. Previous and sporadic seeing measurements were performed on this site but with inconclusive results \cite{per03,dal04,mez13,ric19}.

\section{Observations and Reductions}
The observations were made using the 0.5m OAUNI telescope \cite{per15} during the 2016 and 2017 observational campaigns using the observational windows available in the supernovae (SNe) photometry program (see for example, \cite{per16a,per17a}). We chosen this program for our seeing statistics because consecutive measurements in \textit{V} (effective wavelength, $\lambda_{eff}$ = 551nm) and \textit{R} ($\lambda_{eff}$ = 658nm) broadband filters are usual for a typical supernova target. This fact lets a quasi simultaneous comparison of the seeing conditions in both bands. In general, we observed during the dry season at the central peruvian Andes covering mainly July and August months centered in new moon. As a detector was used a front-illuminated CCD STXL-6303E of 3072$\times$2048 pixels$^2$ and 9$\mu$m/pixel. At the \textit{f}/8.2 Cassegrain focus the detector yields a plate scale of 0.45$"$/pixel and a field-of-view (FOV) of 23$'\times$15$'$. All the measurements were gathered with the detector cooled to -10 C to minimize the dark current.

\begin{table}
\caption{\label{fulltab}Seeing measurements from supernovae fields observed during 2016$-$2017 campaigns.}
\begin{center}
\tiny
\begin{tabular}{lcccccccccc}
\br
SN field & \multicolumn{3}{c}{Local time (UT$-$5)} & filter  & \textit{N} & $\chi_{\rm{mean}}$ & \multicolumn{3}{c}{seeing (")} \vspace*{3px}\\
\cline{2-4} \cline{8-11}\\
 &   date  & begin & end &  &  &   &  obs$_{\rm{mean}}$ & cor$_{\rm{mean}}$ & cor$_{\rm{median}}$ & cor$_{\rm{mode}}$ \vspace*{3px} \\ 
(1) & (2) & (3) & (4) & (5) & (6) & (7) & (8) & (9) & (10) & (11) \\
\mr
PSNJ1410   & 2016/01/10 & 04:08 & 04:21  & \textit{V}  &  30  &   1.411 &  2.37 $\pm$ 0.18 & 1.92 $\pm$ 0.15 & 1.90 & 1.86 \\
          &             & 04:23 & 04:36  & \textit{R}  & 30   &   1.359  & 2.46 $\pm$ 0.21 & 2.05 $\pm$ 0.17 & 2.08 & 1.90 \\
%ASSASN16gq & 2016/07/10 & 20:49 & 21:02    &  30  &   1.224 &  2.066 $\pm$ 0.197 & 1.829 $\pm$ 0.168 &  1.778 \\
%ASSASN16gq & 2016/07/10 & 21:03 & 21:16    &  30  &   1.272 &  1.887 $\pm$ 0.190 & 1.633 $\pm$ 0.167 &  1.650 \\
ASASSN16gq & 2016/07/10 & 20:49 & 21:16 & \textit{V}   &  60  &   1.248 &  2.01 $\pm$ 0.20 & 1.76 $\pm$ 0.18 &  1.75 & 1.69 \\
           &            & 21:21 & 21:35 & \textit{R}   &  30  &   1.342 &  1.91 $\pm$ 0.17 & 1.60 $\pm$ 0.14 & 1.62  & 1.62 \\
ASASSN16gp & 2016/07/11 & 22:01 & 22:27 & \textit{V}   &  60  &   2.647 &  3.12 $\pm$ 0.28 & 1.74 $\pm$ 0.16 & 1.72 & 1.64 \\
           &            & 22:37 & 23:03 & \textit{R}   &  60  &   3.407 &  3.47 $\pm$ 0.40 & 1.66 $\pm$ 0.19 & 1.62 & 1.52 \\        
ASASSN16fp & 2016/07/13 & 03:58 & 04:11 & \textit{V}   &  30  &   1.483 &  2.08 $\pm$ 0.15 & 1.64 $\pm$ 0.13 & 1.62 & 1.55 \\
           &            & 04:13 & 04:26 & \textit{R}   &  30  &   1.571 &  2.05 $\pm$ 0.13 & 1.57 $\pm$ 0.10 & 1.57 & 1.57 \\ 
ASASSN16hw & 2016/08/05 & 21:53 & 22:13 & \textit{V}   &  45  &   1.218 &  2.38 $\pm$ 0.22 & 2.12 $\pm$ 0.19 & 2.08 & 1.99 \\
           &            & 21:31 & 21:51 & \textit{R}   &  45  &   1.282 &  2.46 $\pm$ 0.29 & 2.12 $\pm$ 0.25 & 2.05 & 1.96 \\ 
ASASSN16hz & 2016/08/09 & 04:33 & 04:53 & \textit{V}   &  45  &   1.407 &  2.06 $\pm$ 0.24 & 1.68 $\pm$ 0.21 &  1.69 & 1.72 \\ 
           &            & 04:11 & 04:32 & \textit{R}   &  45  &   1.295 &  1.92 $\pm$ 0.22 & 1.64 $\pm$ 0.19 &  1.62 & 1.65 \\ 
%ASSASN16jt & 2016/09/04 & 20:18 & 20:38    &  45  &   1.389 &  2.395 $\pm$ 0.226 & 1.966 $\pm$ 0.181 &  1.938 \\
%ASSASN16jt & 2016/09/04	& 21:23 & 21:43    &  45  &   1.210 &  2.159 $\pm$ 0.211 & 1.925 $\pm$ 0.187 & 	1.809 \\
ASASSN16jt & 2016/09/04	& 20:18 & 21:43 & \textit{V}     &  90  &   1.285 &  2.32 $\pm$ 0.24 & 2.00 $\pm$ 0.18 & 1.96 & 1.91 \\
           &            & 20:40 & 22:04 & \textit{R}     &  90  &   1.247 &  2.21 $\pm$ 0.18 & 1.94 $\pm$ 0.16 & 1.93 & 1.89 \\ 
SN2017erp  & 2017/07/22	& 20:55 & 21:15 & \textit{V}     &  45  &   1.141 &  2.52 $\pm$ 0.20 & 2.33 $\pm$ 0.18 & 2.32 & 2.22 \\
           &            & 20:09 & 20:29 & \textit{R}     &  45  &   1.049 &  2.86 $\pm$ 0.17 & 2.78 $\pm$ 0.16 & 2.78 & 2.86 \\ 
SN2017erp  & 2017/07/23	& 20:25 & 20:44  & \textit{V}    &  45  &   1.081 &  2.83 $\pm$ 0.43 & 2.70 $\pm$ 0.42 & 2.59 & 2.36 \\
           &            & 20:03 & 20:22  & \textit{R}    &  45  &   1.045 &  2.84 $\pm$ 0.29 & 2.76 $\pm$ 0.28 & 2.73 & 2.72 \\ 
%SN2017erp  & 2017/07/24	& 20:34 & 20:54    &  45  &   1.110 &  2.805 $\pm$ 0.276 & 2.636 $\pm$ 0.268 & 	2.613 \\
%SN2017erp  & 2017/07/24	& 21:17 & 21:38    &  45  &   1.241 &  2.849 $\pm$ 0.269 & 2.504 $\pm$ 0.245 & 	2.206 \\
SN2017erp  & 2017/07/24	& 20:34 & 21:38 & \textit{V}     &  90  &   1.175 &  2.86 $\pm$ 0.27 & 2.60 $\pm$ 0.26 & 2.61 & 2.62 \\
           &            & 20:13 & 21:15 & \textit{R}     &  90  &   1.119 &  2.68 $\pm$ 0.19 & 2.51 $\pm$ 0.19 & 2.50 & 2.51 \\ 
%SN2017erp  & 2017/07/25	& 22:54 & 23:15    &  45  &   2.060 &  2.221 $\pm$ 0.219 & 1.443 $\pm$ 0.144 & 	1.362 \\
%SN2017erp  & 2017/07/25	& 23:38 & 23:59    &  45  &   3.175 &  2.782 $\pm$ 0.277 & 1.392 $\pm$ 0.130 & 	1.299 \\
SN2017erp  & 2017/07/25	& 22:54 & 23:59 & \textit{V}    &  90  &   2.536 &  2.50 $\pm$ 0.37 & 1.44 $\pm$ 0.14 & 	1.43 & 1.55 \\
           &              & 22:32 & 23:37 & \textit{R}    &  90  &   2.136 &  2.25 $\pm$ 0.22 & 1.44 $\pm$ 0.11 &  1.45 & 1.51 \\ 
SN2017erp  & 2017/07/26 & 20:01 & 20:28 & \textit{V}     &  60  &   1.066   &  2.00 $\pm$ 0.19 & 1.92 $\pm$ 0.18 &  1.92 & 1.91 \\
           &            & 19:38 & 19:58 & \textit{R}     &  45  &   1.031   &  2.03 $\pm$ 0.40 & 1.99 $\pm$ 0.39 &  1.91 & 1.73 \\ 
SN2017erp  & 2017/07/27	& 19:50 & 20:10 & \textit{V}     &  45  &   1.050 &  1.98 $\pm$ 0.36 & 1.92 $\pm$ 0.35 & 	1.82 & 1.70 \\
           &              & 19:29 & 19:49 & \textit{R}     &  45  &   1.024 &  2.10 $\pm$ 0.31 & 2.07 $\pm$ 0.30 & 2.07 &  2.08 \\ 
SN2017erp  & 2017/07/28	& 20:07 & 20:27	 & \textit{V}  &  45  &   1.085 &  2.00 $\pm$ 0.17 & 1.90 $\pm$ 0.16 & 1.91 & 1.90 \\
           &              & 19:46 & 20:06  &  \textit{R} &  45  &   1.049 &  2.03 $\pm$ 0.18 & 1.97 $\pm$ 0.18 & 1.97 & 1.96 \\ 
SN2017erp  & 2017/07/29	& 19:54 & 20:14	& \textit{V}     &  45  &   1.068 &  1.86 $\pm$ 0.22 & 1.79 $\pm$ 0.21 & 1.78 & 1.59 \\
           &            & 19:32 & 19:52 & \textit{R}     &  45  &   1.037 &  2.01 $\pm$ 0.22 & 1.97 $\pm$ 0.21 & 1.99 & 1.97 \\ 
SN2017erp  & 2017/08/20	& 19:49 & 20:09 & \textit{V}     &  45  &   1.315 &  2.29 $\pm$ 0.22 & 1.95 $\pm$ 0.17 & 1.94 & 1.78 \\
           &            & 19:26	& 19:46 & \textit{R}     &  45  &   1.221 &  1.80 $\pm$ 0.18 & 1.60 $\pm$ 0.15 & 1.63  & 1.64 \\ 
SN2017erp  & 2017/08/21	& 19:58 & 20:35	 & \textit{V}    &  80  &   1.436 &  2.22 $\pm$ 0.26 & 1.79 $\pm$ 0.20 & 1.77 & 1.71 \\
           &            & 19:27	& 19:56  & \textit{R}    &  60  &   1.259 &  2.15 $\pm$ 0.26 & 1.87 $\pm$ 0.24 & 1.86 & 1.87 \\ 
SN2017erp  & 2017/08/22	& 20:56 & 21:24	 & \textit{V}    &  60  &   1.995 &  2.53 $\pm$ 0.15 & 1.68 $\pm$ 0.10 &  1.67 & 1.62 \\
           &            & 20:27	& 20:54  & \textit{R}    &  60  &   1.655 &  2.49 $\pm$ 0.28 & 1.85 $\pm$ 0.21 &  1.83 & 1.70 \\ 
\br
\end{tabular}
\end{center}
\end{table}

All the images were reduced using \textit{IRAF} \cite{tod86} with typical calibration corrections of dark current and flatfield. Special \textit{IRAF} routines were used to align the sequences of images along with standard routines as \texttt{xregister} and \texttt{imalign}. In order to compute the seeing we measured automatically the \textit{fwhm} of stellar profiles  in a particular supernova sequence in each filter and date. Typical images sequence  consisted of series between 30 and 90 individual frames always with a integration time of twenty seconds (by frame). Special care was taken with frames affected by tracking problems or by occasional vibration of gust of wind. In these cases, the frames were exclude on the following analysis. The mean \textit{fwhm} for the same three brightest stars in each sequence was computed using the \textit{IRAF} routine \texttt{imexamine}. Comparison with similar routines as \texttt{daoedit} \cite{dav94} and \texttt{psfmeasure} \cite{val94a,val94b} resulted in similar values.

The observation log from the supernovae program during the 2016 and 2017 campaigns is shown in Table~\ref{fulltab}. The SN field used for the seeing computation is indicated in Col. (1). During the 2016 campaign were observed seven SN targets in different nights (one by night). On 2017 one target was observed in eleven nights. The local observation date is shown in Col. (2) along with the interval time of the observational windows (Cols. 3$-$4). The last ones are typically between thirteen and forty minutes. The used filter is indicated in Col. (5) with the number of measurements in a given observation sequence in Col (6). The mean airmass during the sequence is shown in Col. (7). Finally, Cols. 8$-$11 indicate the mean seeing without airmass correction (obs$_{\rm{mean}}$) with its 1-sigma dispersion, the proper value with the correction applied (cor$_{\rm{mean}}$) along with the median (cor$_{\rm{median}}$) and the mode (cor$_{\rm{mode}}$) values, respectively.

The total number of individual seeing measurements showed in this work is about two thousand and must be taken with care in the sense that this sample represents an aleatory survey of the local seeing for the OAUNI site. In general, a typical good observation night includes eight or nine continuous hours and we are sampling short observational windows with no more than  1.5 hours by night (including both filters). Nevertheless, we believe that the measurements presented here are representative and useful to analyze the monthly and interannual seeing feature for this site.

\begin{figure}
\includegraphics[width=16cm]{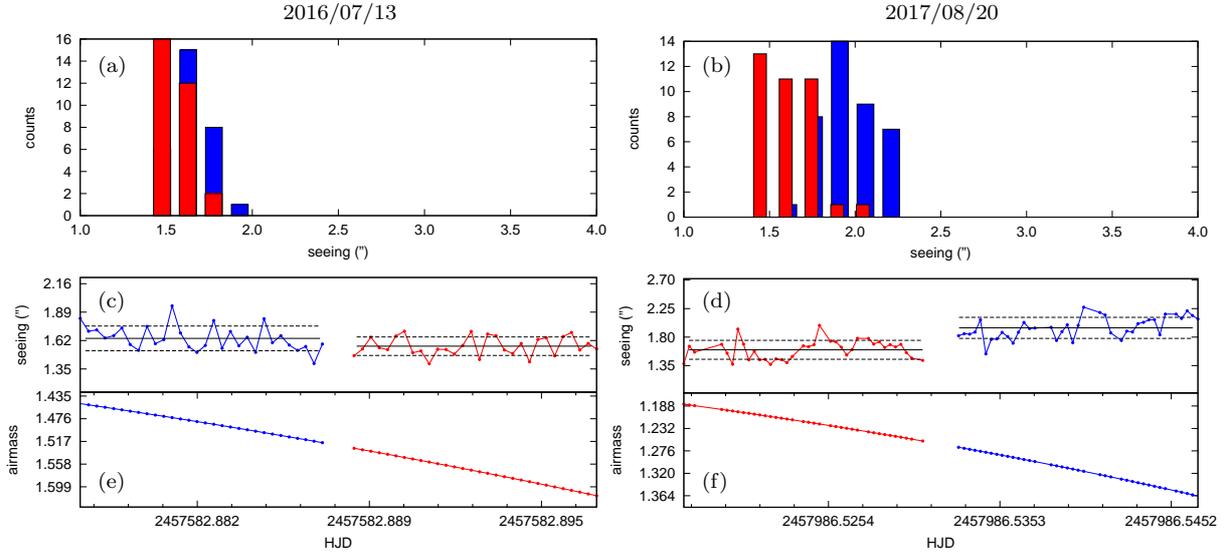}
\caption{\label{gseeing} Good seeing measurements on 2016/07/13 (left) and 2017/08/20 (right). (a,b) Histograms with airmass correction in \textit{V} (blue) and \textit{R} (red) filters. (c,d) Seeing time evolution. Average seeing (solid line) and its 1-sigma dispersion (dotted lines) are indicated for each filter. (e,f) Airmass during the observational window.}
\end{figure}

\begin{figure}
\includegraphics[width=16cm]{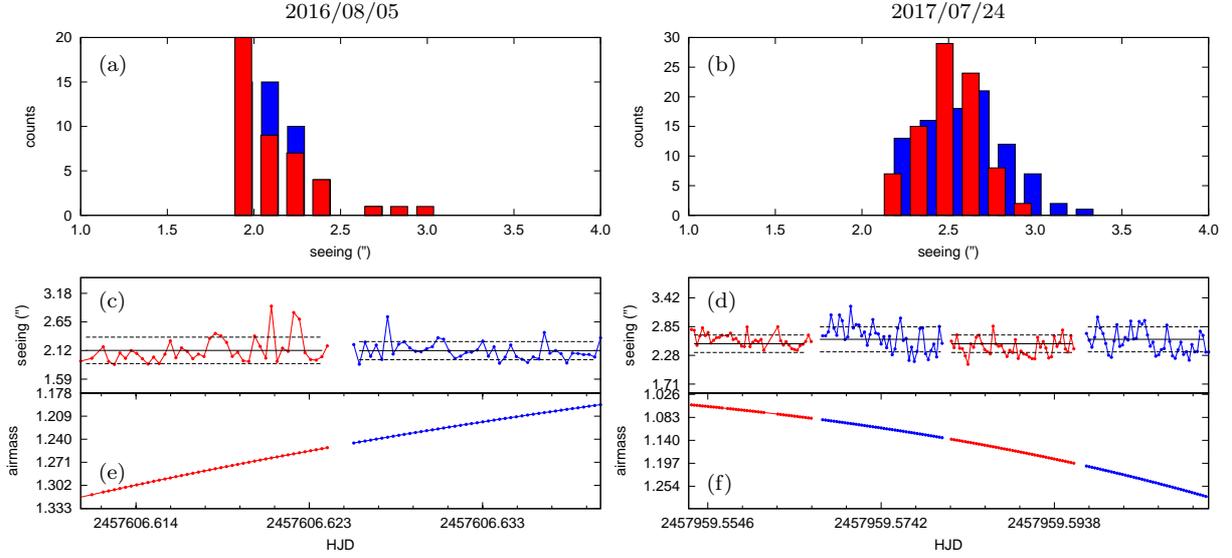}
\caption{\label{bseeing} Bad seeing measurements on 2016/08/05 (left) and 2017/07/24 (right). As Fig.~\ref{gseeing}.}
\end{figure}

\begin{figure}
\includegraphics[width=16cm]{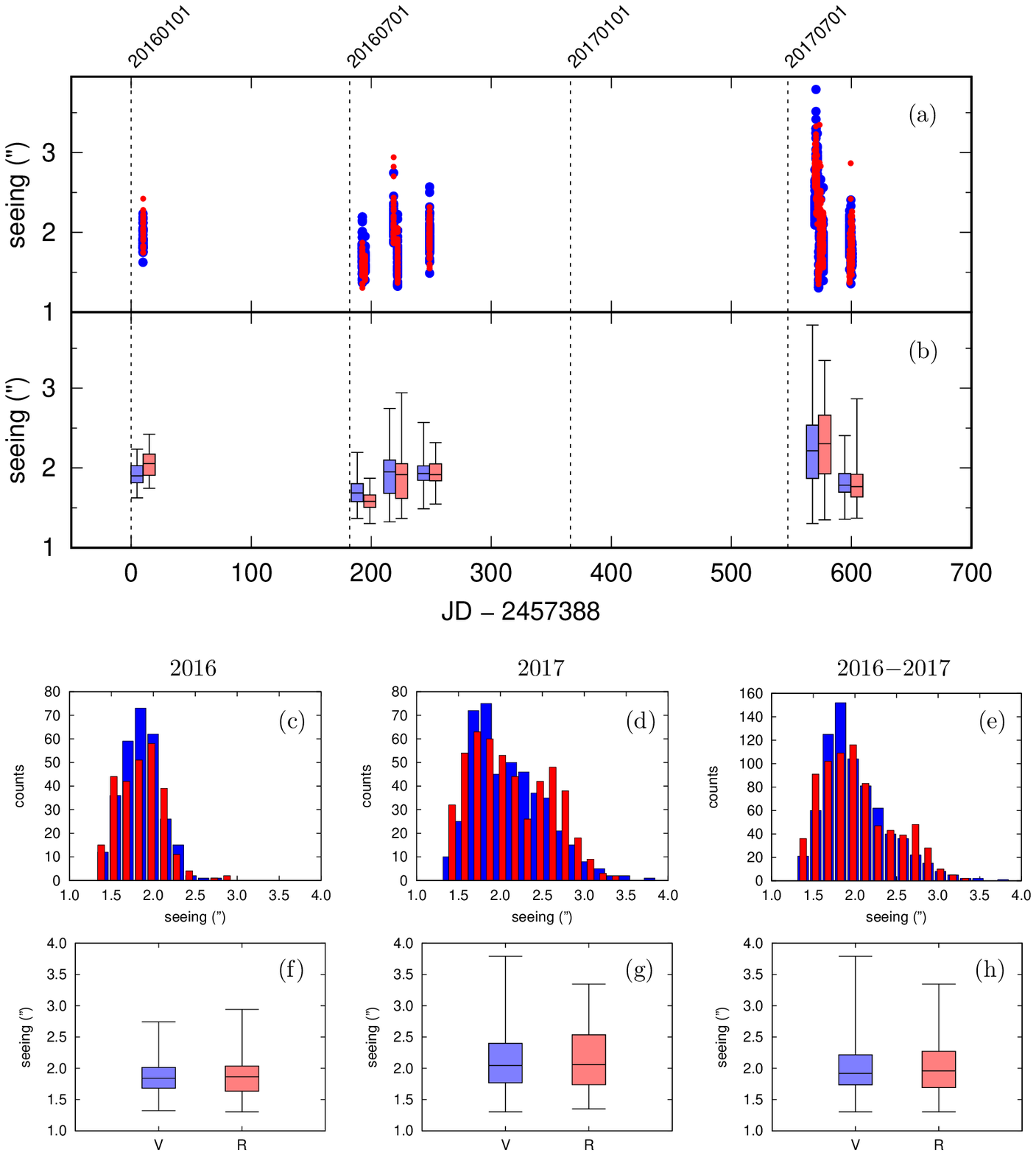}
\caption{\label{seeingtot} (a) Full seeing time evolution with airmass correction between 2016$-$2017. Individual measurements in \textit{V} (blue) and \textit{R} (red) filters.  (b) Monthly seeing analysis for the full sample showing minimum, first quartile, median, third quartile and maximum in each case. (c,d,e) Histograms for 2016, 2017 and full sample. (f,g,h) Interannual seeing analysis.}
\end{figure}

\section{Analysis}

As indicated in \S1, the \textit{fwhm} airmass correction applies for $\chi$ $<$ 2. At least in two cases (2016/07/11 and 2017/07/25 nights) the mean airmasses (see Table \ref{fulltab}) are higher than this limit. In general, higher airmasses underestimated the seeing values. In order to avoid this situation, the following analysis considers only seeing values under this limit resulting in about five hundred measurements filtered of the full sample.

Figures \ref{gseeing} and \ref{bseeing} show examples of good and bad seeing nights, respectively. The observational windows in Fig.~\ref{gseeing}-left for 2016/07/13 night indicate better seeing conditions with a median as low as 1$\farcs$62 and 1$\farcs$57 for \textit{V} and \textit{R} bands, respectively (see Table~\ref{fulltab}). The sampled time interval was of 30 min with $\chi$ $<$ 1.6 observing to the West. In general, lower seeing values in \textit{R} than \textit{V} are expected under normal conditions (\S1). The same pattern is observed on 2017/08/20 night (Fig.~\ref{gseeing}-right) with a median seeing of 1$\farcs$63 in \textit{R} band but the seeing being worst in \textit{V} (1$\farcs$94). In this case, the monitoring was of $\sim$45 min with $\chi$ $<$ 1.4 also looking to the West. Considering the mean values of these nights, typical 1-sigma dispersion lower than $\sim$0$\farcs$15 is indication of a stable observational window. On the other hand, the 2016/08/05 night in Fig.~\ref{bseeing}-left show bad seeing conditions in a 42 min observational window with typical median seeing of 2$\farcs$1 in both filters and with a little bit larger 1-sigma dispersion ($\sim$0$\farcs$22). In this case, the observations were made with  $\chi$ $<$ 1.3 on the East hemisphere. Worst seeing conditions were found on 2017/07/24 night (Fig.~\ref{bseeing}-right). Two alternative sequences in both filters were acquired during a 90 min observational window with a higher median seeing ($>$ 2$\farcs$5) and 1-sigma dispersion ($\sim$ 0$\farcs$23). In this night the West hemisphere was monitored with $\chi$ $<$ 1.2. Considering the last two cases, a 1-sigma dispersion of 0$\farcs$22 can still characterize stable conditions, even for nights with bad seeing. It is interesting because the photometry for several science programs requires primarily a stable atmosphere even although the intrinsic seeing is not the best. Finally, 83\% of the analyzed nights on 2016 shown median seeing values following the expected wavelength dependence, i.e. better and smaller seeing in \textit{R} filter as compared with the \textit{V} values. Nevertheless, on 2017 this pattern only is seen in 30\% of the nights and it seems indicate an anomalous trend.

\begin{table}
\caption{\label{tabmonth} Monthly seeing analysis.}
\begin{center}
\tiny
\begin{tabular}{lccccccc}
\br
year &  \textit{N} & filter & \multicolumn{5}{c}{seeing (")} \vspace*{2px}\\
\cline{4-8}
      &       &   &  \rm{mean} & \rm{Q}$_{1}$ & median & \rm{Q}$_{3}$ & \rm{mode}  \\ 
\mr
2016/01      &     30      &  \textit{V}  &   1.92 $\pm$ 0.15  & 1.81  & 1.90  & 2.03 & 1.75   \\
             &     30      &  \textit{R}  &   2.05 $\pm$ 0.17  & 1.91  & 2.05  & 2.17 & 2.10   \\
2016/07      &     90      &  \textit{V}  &   1.70 $\pm$ 0.17  & 1.58  & 1.69  & 1.80 & 1.58   \\
             &     59      &  \textit{R}  &   1.58 $\pm$ 0.12  & 1.51  & 1.58  & 1.66 & 1.53   \\
2016/08      &     90      &  \textit{V}  &   1.90 $\pm$ 0.29  & 1.68  & 1.95  & 2.10 & 1.68   \\
             &     88      &  \textit{R}  &   1.88 $\pm$ 0.32  & 1.62  & 1.92  & 2.05 & 1.98   \\                                                     
2016/09      &     77      &  \textit{V}  &   1.94 $\pm$ 0.18  & 1.84  & 1.93  & 2.02 & 1.86   \\
             &     90      &  \textit{R}  &   1.94 $\pm$ 0.16  & 1.84  & 1.92  & 2.05 & 1.83   \\                                         
2017/07      &     323     &  \textit{V}  &   2.23 $\pm$ 0.45  & 1.87  & 2.21  & 2.54 & 2.20   \\ 
             &     346     &  \textit{R}  &   2.27 $\pm$ 0.46  & 1.92  & 2.30  & 2.66 & 2.68   \\
2017/08      &     126     &  \textit{V}  &   1.82 $\pm$ 0.19  & 1.69  & 1.78  & 1.93 & 1.72   \\
             &     146     &  \textit{R}  &   1.79 $\pm$ 0.23  & 1.64  & 1.77  & 1.92 & 1.68   \\                                                    
\br
\end{tabular}
\end{center}
\end{table}

\begin{table}
\caption{\label{tabyear} Annual analysis for seeing .}
\begin{center}
\tiny
\begin{tabular}{lccccccc}
\br
year &  \textit{N} & filter & \multicolumn{5}{c}{seeing (")} \vspace*{2px}\\
\cline{4-8}
      &       &   &  \rm{mean} & \rm{Q}$_{1}$ & median & \rm{Q}$_{3}$ & \rm{mode} \\ 
\mr
2016    &     287     &  \textit{V}  &   1.85 $\pm$ 0.24  & 1.68  & 1.84  & 2.01 & 1.79  \\ 
         &     267     &  \textit{R}  &   1.85 $\pm$ 0.27  & 1.63  & 1.86  & 2.04 & 2.04  \\
2017     &     449     &  \textit{V}  &   2.11 $\pm$ 0.43  & 1.77  & 2.04  & 2.40 & 1.72  \\
         &     492     &  \textit{R}  &   2.13 $\pm$ 0.46  & 1.74  & 2.06  & 2.54 & 1.92  \\
2016-17  &     736     &  \textit{V}  &   2.01 $\pm$ 0.39  & 1.74  & 1.92  & 2.21 & 1.86  \\
         &     759     &  \textit{R}  &   2.03 $\pm$ 0.43  & 1.69  & 1.96  & 2.27 & 1.87 \\
\br
\end{tabular}
\end{center}
\end{table}

In order to help the interannual seeing comparison and/or detect any seasonal pattern, Fig.~\ref{seeingtot} shows all the measurements collected on 2016 and 2017 campaigns in both filters. The full seeing evolution considering the individual mesurements is indicated in Fig.~\ref{seeingtot}a and the monthly analysis showing the basic statistics is shown in Fig.~\ref{seeingtot}b and Table~\ref{tabmonth}. As we can see, 2016 was sampled in four months (January, July, August and September) and 2017 only in two (July and August). Nevertheless, the number of  measurements on 2016 (554) was a little bit higher than the half ones on 2017 (941). Therefore, our sample let to compare the months of July and August in both years. In general, the best seeing conditions are obtained on 2016 July with a median of 1$\farcs$6 and 1$\farcs$7 on \textit{R} and \textit{V} band, respectively. On the other hand, 2017 July presented the worst seeing values as compared with the rest of months, with a median seeing above of 2$\farcs$2. Interestingly, 2016 August and 2017 August show similar mean values (1$\farcs$9 and 1$\farcs$8, respectively) in both filters. Considering the dry season at the central peruvian Andes (between May and August), the best seeing conditions are expected to happen on July. In this sense our results for 2017 seems to indicate an anomalous year which could be affected by the extreme climatic event Ni\~{n}o Costero just ended on 2017 May. This is reinforced by the fact that the clear sky rate measured in our site by ground-based weather radar \cite{mar20} drops from 84\% in 2016 July to 77\% on 2017 July and. The poor seeing wavelength dependence on 2017 appointed above also corroborates this finding. However, the normal conditions on 2016 seem to indicate that good seeing can be achieved at a level of 1$\farcs$5 if we considering the first quartile founded on July of that year. For the sake of completeness, we noted that the occasional open nights at the beginning (2016 September) and in the middle (2016 January) of a typical rainy season on our site shown moderate seeing values ($\sim$1$\farcs$9$-$2$\farcs$0).

Finally, the interannual seeing comparison is shown in Figs.~\ref{seeingtot}c$-$e with histograms for 2016, 2017 and both years, respectively. The statistics analysis is indicated in Figs.~\ref{seeingtot}f$-$g and Table~\ref{tabyear}. The \textit{V} and \textit{R} mean values are similar in all cases considering their typical dispersion. The 2016 sample is characterized by a median seeing of 1$\farcs$84 and 1$\farcs$86 at \textit{V} and \textit{R} bands, respectively. On the other hand, the 2017 sample shows a worst seeing with a median of 2$\farcs$04 on \textit{V} and 2$\farcs$06 on \textit{R} and reflecting the irregular conditions of 2017 July. The full sample considering the measurements combined by the  two years has a median seeing of 1$\farcs$92 and 1$\farcs$96 for \textit{V} and \textit{R} bands, respectively. The statistics shown here indicate that our site has a reasonable sky quality for astronomical observations. As comparison, the best sites in the world with astronomical facilities have median seeing values lower than 1$\farcs$0 \cite{mic03}.

\section{Conclusions}

This work shows seeing measurements obtained at OAUNI site during 2016 and 2017 observational campaigns. This is part of an ongoing site testing program at the central peruvian Andes. The full-width-at-half-maximum of stellar profiles were measured directly on astronomical images of our supernovae scientific program yielding around one thousand and a half useful individual measurements. The seeing of a total of eighteen nights was evaluated in two filters during both campaigns. The comparative monthly analysis indicates better observational conditions for optical astronomical during the dry season on the site. This was particularly true on 2016 with a typical median seeing of 1$\farcs$8. On the other hand, 2017 shown to be an anomalous year with a worst median seeing of 2$\farcs$0 and far for showing a typical seeing wavelength dependence. The extreme climate phenomena of the first semester on that year can explain this particular feature. In general, the best seeing conditions considering both years were present on 2016 July with the best monthly quartile of 1$\farcs$5. These findings suggest a reasonable sky quality for optical astronomical observations at OAUNI site. In this sense, it is appropriate to indicate that the seeing levels reported here do not prevent the development of selected astronomical scientific programs \cite{per20a,per20b,per20c} . 

\ack
The author is grateful for the economic support from The World Academy of Sciences (TWAS), Rectorate and the Instituto General de Investigaci\'on (IGI) at UNI, and Concytec (Convenio 102-2015 Fondecyt). Special thanks to the Huancayo Observatory staff for the logistic support.

\end{document}